\documentclass[journal]{IEEEtran}
\usepackage{cite}% *** CITATION PACKAGES ***
\usepackage[cmex10]{amsmath}% *** MATH PACKAGES ***
\usepackage{amssymb}
\usepackage{amsthm}
\usepackage{mathrsfs}%SF
\usepackage{bm}
\usepackage[mathscr]{eucal}
\usepackage{amssymb,amsmath,amsthm,amsfonts,latexsym}
\usepackage{amsmath,graphicx,bm,xcolor,url}
\usepackage{graphicx}
\usepackage{latexsym}%SF
\usepackage{CJK}%SF
\usepackage{indentfirst}%SF
\usepackage{psfrag}%SF
\usepackage{setspace}%SF
\usepackage{algorithmic}% *** SPECIALIZED LIST PACKAGES ***
\usepackage{algorithmic, cite}
\usepackage{algorithm}
\usepackage{array}% *** ALIGNMENT PACKAGES ***
\usepackage{mdwmath}
\usepackage{mdwtab}
\usepackage{eqparbox}
\usepackage{url}% *** PDF, URL AND HYPERLINK PACKAGES ***
\usepackage{epstopdf}
\usepackage{epsfig,epsf,psfrag}
\usepackage{fixltx2e}%ordering of single and double column floats
\usepackage{verbatim}
\usepackage{textcomp}
\usepackage{psfrag} %%25 June

\usepackage{caption}
\theoremstyle{plain}

\theoremstyle{remark}
\newtheorem{mylem}{Lemma}
\theoremstyle{plain}

\theoremstyle{remark}

\theoremstyle{plain}

\theoremstyle{remark}

\theoremstyle{remark}

\theoremstyle{remark}

\theoremstyle{remark}

\theoremstyle{remark}

\theoremstyle{remark}

\catcode`~=11 \def\UrlSpecials{\do\~{\kern -.15em\lower .7ex\hbox{~}\kern .04em}} \catcode`~=13

\allowdisplaybreaks[4]

\newcommand{\calA}{\mathcal{A}}

\newcommand{\calC}{\mathcal{C}}

\newcommand{\calN}{\mathcal{N}}

\newcommand{\calQ}{\mathcal{Q}}

\newcommand{\ba}{\mathbf{a}}
\newcommand{\bA}{\mathbf{A}}

\newcommand{\bc}{\mathbf{c}}
\newcommand{\bC}{\mathbf{C}}

\newcommand{\boldf}{\mathbf{f}}

\newcommand{\bg}{\mathbf{g}}
\newcommand{\bG}{\mathbf{G}}
\newcommand{\bh}{\mathbf{h}}
\newcommand{\bH}{\mathbf{H}}

\newcommand{\bI}{\mathbf{I}}

\newcommand{\bM}{\mathbf{M}}

\newcommand{\bT}{\mathbf{T}}
\newcommand{\bu}{\mathbf{u}}

\newcommand{\bW}{\mathbf{W}}

\newcommand{\by}{\mathbf{y}}

\newcommand{\bz}{\mathbf{z}}

\newcommand{\rmd}{\mathrm{d}}

\newcommand{\rmH}{\mathrm{H}}

\newcommand{\rmM}{\mathrm{M}}

\newcommand{\rmp}{\mathrm{p}}

\newcommand{\rmt}{\mathrm{t}}
\newcommand{\rmT}{\mathrm{T}}

\newcommand{\bbC}{\mathbb{C}}

\newcommand{\bbE}{\mathbb{E}}

\newcommand{\bbV}{\mathbb{V}}

\DeclareMathAlphabet{\mathbsf}{OT1}{cmss}{bx}{n}
\DeclareMathAlphabet{\mathssf}{OT1}{cmss}{m}{sl}% slanted sans serif

\DeclareSymbolFont{bsfletters}{OT1}{cmss}{bx}{n}
\DeclareSymbolFont{ssfletters}{OT1}{cmss}{m}{n}
\DeclareMathSymbol{\bsfGamma}{0}{bsfletters}{'000}
\DeclareMathSymbol{\ssfGamma}{0}{ssfletters}{'000}
\DeclareMathSymbol{\bsfDelta}{0}{bsfletters}{'001}
\DeclareMathSymbol{\ssfDelta}{0}{ssfletters}{'001}
\DeclareMathSymbol{\bsfTheta}{0}{bsfletters}{'002}
\DeclareMathSymbol{\ssfTheta}{0}{ssfletters}{'002}
\DeclareMathSymbol{\bsfLambda}{0}{bsfletters}{'003}
\DeclareMathSymbol{\ssfLambda}{0}{ssfletters}{'003}
\DeclareMathSymbol{\bsfXi}{0}{bsfletters}{'004}
\DeclareMathSymbol{\ssfXi}{0}{ssfletters}{'004}
\DeclareMathSymbol{\bsfPi}{0}{bsfletters}{'005}
\DeclareMathSymbol{\ssfPi}{0}{ssfletters}{'005}
\DeclareMathSymbol{\bsfSigma}{0}{bsfletters}{'006}
\DeclareMathSymbol{\ssfSigma}{0}{ssfletters}{'006}
\DeclareMathSymbol{\bsfUpsilon}{0}{bsfletters}{'007}
\DeclareMathSymbol{\ssfUpsilon}{0}{ssfletters}{'007}
\DeclareMathSymbol{\bsfPhi}{0}{bsfletters}{'010}
\DeclareMathSymbol{\ssfPhi}{0}{ssfletters}{'010}
\DeclareMathSymbol{\bsfPsi}{0}{bsfletters}{'011}
\DeclareMathSymbol{\ssfPsi}{0}{ssfletters}{'011}
\DeclareMathSymbol{\bsfOmega}{0}{bsfletters}{'012}
\DeclareMathSymbol{\ssfOmega}{0}{ssfletters}{'012}

\newcommand{\hatbc}{\widehat{\bc}}

\newcommand{\tilbu}{\widetilde{\bu}}

\newcommand{\tilby}{\widetilde{\by}}

\newcommand{\hatbz}{\widehat{\bz}}

\newcommand{\balpha}{\bm{\alpha}}

\newcommand{\bLambda}{\bm{\Lambda}}
\newcommand{\bSigma	}{\bm{\Sigma}}

\def\norm#1{\left\| #1 \right\|}
\def\norm2#1{\left\| #1 \right\|_2}
\def\norm22#1{\left\| #1 \right\|_2^2}

\DeclareMathOperator{\diag}{diag}

\DeclareMathOperator{\tr}{tr}

\newcommand{\qednew}{\nobreak \ifvmode \relax \else
      \ifdim\lastskip<1.5em \hskip-\lastskip
      \hskip1.5em plus0em minus0.5em \fi \nobreak
      \vrule height0.75em width0.5em depth0.25em\fi}

\usepackage{float}
\usepackage{cite}
\usepackage{times,amsmath,color,amssymb,graphicx,epsfig,multirow,float,algorithm,algorithmic}
\usepackage{amsthm}
\usepackage{caption}
\usepackage{subcaption}
\usepackage[utf8]{inputenc}
\newtheorem{remark}{Remark}
\usepackage[hidelinks]{hyperref}
\begin{document}
\captionsetup{font={small}}

\title{Multi-Antenna Broadband Backscatter Communications}

% \author{xxx}

\author{Hao Chen, Zhizhi Huang, Ying-Chang Liang, \IEEEmembership{Fellow,~IEEE}, and Robert Schober, \IEEEmembership{Fellow,~IEEE}
%	\thanks{Part of this work was presented in IEEE GLOBECOM'22 \cite{conference}. This work was supported in part by the National Key Research and Development Program of China under Grant 2018YFB1801105; in part by the Key Areas of Research and Development Program of Guangdong Province, China, under Grant 2018B010114001; in part by the Fundamental Research Funds for the Central Universities under Grant ZYGX2019Z022; and in part by the Program of Introducing Talents of Discipline to Universities under Grant B20064. \emph{(Corresponding author: Ying-Chang Liang.)}}
    \thanks{This work has been submitted to the IEEE for possible publication.
Copyright may be transferred without notice, after which this version may
no longer be accessible.}
	\thanks{H.~Chen and Z. Huang are with the National Key Laboratory of Wireless Communications, and the Center for Intelligent Networking and Communications (CINC), University of Electronic Science and Technology of China (UESTC), Chengdu 611731, China (e-mail: {hhhaochen@std.uestc.edu.cn;~zhizhihuang\_uestc@163.com}).}
 % , and also with the Yangtze Delta Region Institute (Huzhou), University of Electronic Science and Technology of China, Huzhou 313001, P. R. China
%	\thanks{Q.~Zhang and R.~Long are with the National Key Laboratory of Wireless Communications, University of Electronic Science and Technology of China, Chengdu 611731, P. R. China (e-mail: {qqzhang\_kite@163.com;~ruizhelong@gmail.com}).}
	\thanks{Y.-C.~Liang is with the Center for Intelligent Networking and Communications (CINC), University of Electronic Science and Technology of China (UESTC), Chengdu 611731, China (e-mail: {liangyc@ieee.org}).}
 \thanks{R. Schober is with the Institute for Digital Communications, Friedrich-Alexander-Universit\"at (FAU) Erlangen-N\"urnberg, 91058 Erlangen, Germany (e-mail: {robert.schober@fau.de}).}
 % , and also with the Yangtze Delta Region Institute (Huzhou), University of Electronic Science and Technology of China, Huzhou 313001, P. R. China
%	\thanks{Y. Pei is with the Singapore Institute of Technology, 138683, Singapore (e-mail: yiyang.pei@singaporetech.edu.sg).}
}
\maketitle
\begin{abstract}
Backscatter communication offers a promising solution to connect massive Internet-of-Things (IoT) devices with low cost and high energy efficiency. Nevertheless, its inherently passive nature limits transmission reliability, thereby hindering improvements in communication range and data rate. To overcome these challenges, we introduce a bistatic broadband backscatter communication (BBBC) system, which equips the backscatter device (BD) with multiple antennas. In the proposed BBBC system, a radio frequency (RF) source directs a sinusoidal signal to the BD, facilitating single-carrier block transmission at the BD. Meanwhile, without requiring channel state information (CSI), cyclic delay diversity (CDD) is employed at the multi-antenna BD to enhance transmission reliability through additional cyclically delayed backscattered signals. We also propose a receiver design that includes preprocessing of the time-domain received signal, pilot-based parameter estimation, and frequency-domain equalization, enabling low-complexity detection of the backscattered signal. Leveraging the matched filter bound (MFB), we analyze the achievable diversity gains in terms of outage probability. Our analysis reveals that spatial diversity is achievable under general Rayleigh fading conditions, and both frequency and spatial diversity are attainable in scenarios where the forward link experiences a line-of-sight (LoS) channel.
Simulation results validate the effectiveness of the proposed BBBC system.
As the number of BD antennas increases, our results show that the proposed scheme not only enhances array gain but also improves diversity order, significantly reducing both outage probability and bit error rate (BER). Consequently, it outperforms conventional schemes that yield only minor gains. 
% , which significantly outperforms conventional schemes, especially as the number of antennas increases.
\end{abstract}


\section{Introduction}
Ambient Internet of Things (AIoT), a network of devices that harness energy from ambient power sources to facilitate communication, aims to achieve sustainable connectivity with low operational costs and high energy efficiency in the forthcoming sixth-generation (6G) systems \cite{6GSCIS, AIoT, 3gppAIoT}. First introduced in 1948 \cite{FirstPropose}, backscatter communication has since been a focal point for research and development in both academia and industry for decades, playing an essential role in the evolution of AIoT \cite{PIEEEBackscatter, RFID, JCINbackscatter}. In backscatter communications, IoT tags, also known as backscatter devices (BDs), reflect and modulate incident radio frequency (RF) signals from ambient sources. By modifying the amplitude and phase of incident signals, BDs embed information directly into them, thereby eliminating the need for power-hungry RF components and even internal energy sources, which results in low manufacturing costs and environmental sustainability of BDs.

A typical backscatter communication system consists of three components: an RF source that emits RF signals, a BD that modulates and conveys information, and a reader that receives the backscattered signal. In particular, there exist three types of system architectures: monostatic backscatter communication (MBC) \cite{FirstPropose, RFID, GainsMBC}, bistatic backscatter communication (BBC) \cite{increasedrangeBBC, NovalInterferenceMitigation, MIMOBBC}, and ambient backscatter communication (ABC) \cite{ABCfirstpropose, modulationintheair, CABC}. In MBC, the RF source and the reader are integrated into a single device, leading to a significant round-trip path loss of the backscattered signal \cite{GainsMBC}. 
% When the RF source and the reader share the same RF component, full duplex technology is necessary to enable simultaneous transmission and reception. This architecture is commonly utilized in Radio Frequency IDentification systems \cite{RFIDoverview}. 
On the contrary, BBC separates the RF source from the reader. With proper deployment, BBC can reduce the path loss of the backscattered signal, thereby extending its communication range \cite{increasedrangeBBC}. Both the MBC and BBC architectures require a dedicated RF source to provide a carrier signal to the BD for transmission. Similarly, the ABC architecture adopts a bistatic configuration where the RF source and the reader are separate. However, in ABC, the BD uses ambient modulated signals such as TV and Wi-Fi signals as carriers \cite{ABCfirstpropose}, thereby eliminating the need for a dedicated RF source. This architecture, however, introduces significant challenges in interference mitigation from these unknown ambient signals \cite{modulationintheair, CABC}.

There exists a growing demand for extended communication range and increased data rate in backscatter communications \cite{AIoT, PIEEEA}. To meet these demands, conventional methods mainly focus on enhancing modulation schemes. On the one hand, besides adopting the BBC architecture as aforementioned, the communication range can be improved by choosing a proper modulation scheme. Frequency-domain modulation, such as frequency-shift keying (FSK) and chirp spread spectrum (CSS), can effectively extend communication range to hundreds or even thousands of meters thanks to their robustness against noise and self-interference \cite{FSK, LoRaBackscatter}. However, these techniques typically lead to a reduction in data rate. On the other hand, to enhance the data rate, higher-order modulation schemes are employed in backscatter communication. In \cite{QAM}, the authors develop backscatter circuits that modify both the amplitude and phase of the backscattered signal, enabling the transmission of higher-order quadrature amplitude modulation (QAM) signals. Yet, this approach results in decreased demodulation sensitivity as the order of modulation increases. Since the aforementioned methods do not tackle the fundamental issue of limited transmission reliability inherent to the passive nature of backscatter communication, they fail to enhance both communication range and data rate. 
% Moreover, it is proved that multipath fading also exists in backscatter radio systems \cite{MultiPath}, which potentially worsens the performance of backscatter communications with a larger communication range and a higher data rate.

An alternative approach to increase the data rate in backscatter communication is to allocate additional frequency bandwidth, which allows for a higher symbol rate at the BD. However, due to the presence of multipath fading \cite{MultiPath}, higher symbol rates can easily result in frequency-selective propagation of the backscattered signal and thereby inter-symbol interference (ISI) that significantly influences signal reception. By transmitting symbols across orthogonal subcarriers, orthogonal frequency division multiplexing (OFDM) can effectively overcome frequency-selective fading, treating each subcarrier channel as a single-tap channel in the frequency domain. In \cite{OFDMbackscatter}, using IQ backscatter modulators, the authors propose a multicarrier BD that can generate OFDM signals by backscattering a single-tone carrier. 
However, the passive nature of backscatter communications constrains the amplitude of the reflected signal. With the high peak-to-average-power ratio (PAPR) of OFDM signals, the average power of reflected signals can be reduced, thereby limiting the transmission performance.
% However, this approach can introduce additional cost and power consumption for the BD, while facing challenges like high peak-to-average-power ratio (PAPR) and sensitivity to carrier frequency offsets (CFO). 
In addition, leveraging frequency-domain equalization, single-carrier block transmission serves as an effective alternative to deal with the frequency-selective fading channel, avoiding the above bottleneck associated with OFDM \cite{Sari}. Unlike OFDM, which requires a discrete Fourier transform (DFT) module embedded in the transmitter, single-carrier block transmission concentrates signal processing complexity primarily in the receiver \cite{CMSC}, which aligns well with the low-cost requirement of the BD. Moreover, single-carrier block transmission can realize diversity gains in the receiver using a proper equalizer \cite{OFDMorSC}, whereas OFDM requires additional precoding in the transmitter to attain similar benefits \cite{precodingOFDM}.

Furthermore, to address the limited transmission reliability, multiple backscatter antennas can be deployed at the BD to strengthen the weak backscattered link \cite{GainsMBC, MultiPath, PIEEEA}. Utilizing Van Atta Arrays, which consist of antennas connected in symmetrical pairs \cite{VanAttaArray}, the multi-antenna BD can directly focus array beams back towards the RF source in the MBC system \cite{retridirection}. However, such antenna arrays are unsuitable for bistatic scenarios. Instead, by selecting specific backscatter antennas for transmission, spatial modulation (SM) can strengthen the backscattered link and meanwhile simplify the receiver design \cite{AntennaSelection, GSSK, SMMA, RPM, PBIT}. Since only a subset of antennas is utilized in the SM scheme, the gain provided by the antenna array can be degraded. Additionally, optimizing the passive beamforming vector of the BD can greatly increase the array gain \cite{HUZHOU, MacroBase, MIMOSR, HUJINLIN, HUAMENG, Protocol&Beamforming}. However, this optimization raises challenges in acquiring BD-related channel state information (CSI). Alternatively, the multi-antenna BD can employ a space-time code (STC) to achieve transmit diversity without requiring CSI \cite{Spacetimecodebackscatter, BLSTC, BetterASTC, ABCSTC}. Yet, it is challenging to extend the STC method to the scenario with a large number of backscatter antennas. As another low-complexity but efficient transmit diversity technique, cyclic delay diversity (CDD) can be implemented by the multi-antenna BD. Specifically, by cyclically shifting transmitted signals, CDD creates additional delayed signals arriving at the receiver. Hence, this technique enhances the frequency selectivity of the propagation channel and potentially increases the transmit diversity gain provided by the multi-antenna BD \cite{BIGDFE, WEN, DCDD}.

To overcome the above challenges in improving the data rate and extending the communication range for backscatter communications, in this paper, we introduce a broadband BBC (BBBC) system based on single-carrier block transmission and CDD. Specifically, the contributions of this paper are summarized as follows:
\begin{itemize}
    \item We introduce a BBBC system that consists of a single-antenna RF source, a single-antenna reader, and a multi-antenna BD. The RF source emits a sinusoidal signal to the BD. By backscattering the incident sinusoidal signal, the BD transmits its information using single-carrier block transmission. To reduce the power consumption at the BD, we employ a zero-padding (ZP) scheme between adjacent symbol blocks instead of the traditional cyclic prefix (CP). Meanwhile, each antenna of the BD backscatters the signal with a distinct cyclic delay, thereby increasing transmit diversity for the backscattered signal.
	
    \item Based on the proposed transmission design, we introduce a receiver design to effectively retrieve the signal of the BD. First, the direct-link interference is mitigated by averaging the received signal in the time domain. Then, the overlap-add (OLA) method is employed to construct a circulant channel matrix. Based on pilot signals conveyed by the BD, the reader can estimate the cascaded channel and precisely remove the residual interference. Finally, frequency-domain equalization is employed to extract the information conveyed by the BD.

    \item Exploiting the matched filter bound (MFB), we analyze the achievable diversity order of our proposed system in terms of outage probability. 
    Considering that the backward-link channel exhibits frequency-selective fading, we analyze two channel conditions: one where both the forward and backward links follow Rayleigh fading, and another one where the forward-link channel is a line-of-sight (LoS) channel, while the backward-link channel follows Rayleigh fading.
    Our analysis shows that in the first case, only spatial diversity is available, whereas in the second case, both frequency and spatial diversity are attainable.

    \item Finally, simulation results confirm the effectiveness of our proposed system and validate our analysis of outage probability and diversity order. Moreover, our results demonstrate that with the enhanced diversity, the proposed transmission scheme can outperform conventional schemes, particularly when the number of antennas is large.
\end{itemize}

\emph{Organization: }This paper is organized as follows. Section \ref{systemmodelsection} describes the proposed system model with single-carrier block transmission and CDD. Section \ref{receiverdesign} presents the proposed receiver design to recover the information of the BD. Section \ref{diversityanalysis} analyzes the outage probability and the diversity order under different channel conditions. Section \ref{simulationresults} provides simulation results to validate the effectiveness of our proposed design. Finally, Section \ref{conclusion} concludes this paper.

\emph{Notations: }Boldface lowercase letters $\ba$, boldface uppercase letters $\bA$, and calligraphic uppercase letters $\calA$ denote vectors, matrices, and discrete finite sets, respectively. $\bA^b$, $\bA^{-1}$, $\bA^\rmT$, and $\bA^\rmH$ denote the $b$-th power, inverse, transpose, and conjugate transpose of matrix $\bA$, respectively. $\bA(a : b, :)$ and $\bA(:, c : d)$ denotes the submatrices of $\bA$ formed by rows $a$ through $b$ and columns $c$ through $d$, respectively. $\bm{1}_{a}$ and $\bm{0}_b$ denote an all-one column vector of dimension $a$ and an all-zero column vector of dimension $b$, respectively. $\bm{1}_{a \times b}$ and $\bm{0}_{c \times d}$ denote an all-one matrix of dimension $a \times b$ and an all-zero matrix of dimension $c \times d$, respectively. $\bI_a$ denotes an identity matrix of dimension $a\times a$. $\calC\calN(\mu, \bSigma)$ denotes the circularly symmetric complex Gaussian (CSCG) distribution with mean vector $\bm{\mu}$ and covariance matrix $\bSigma$. $\bbE[a]$ denotes the expectation of random variable $a$. $\diag(\ba)$ denotes the diagonal matrix whose diagonal elements are given by $\ba$. $\tr(\mathbf{A})$ denotes the trace of matrix $\mathbf{A}$.

\section{System Model} \label{systemmodelsection}

\begin{figure}[t!]
	\centering    
	\includegraphics[width=0.9\columnwidth]{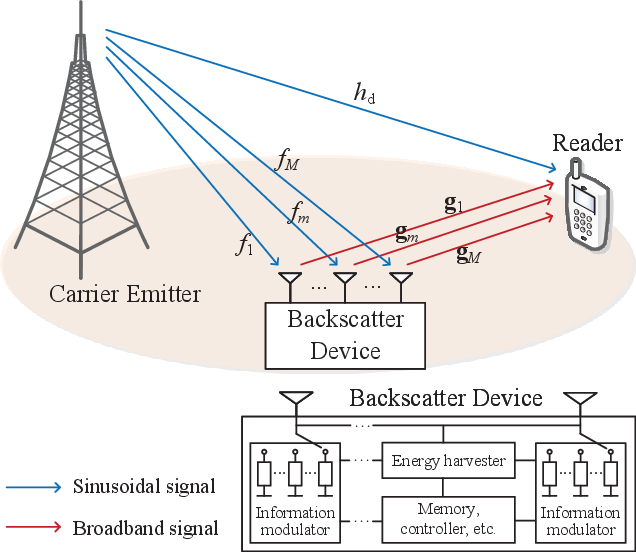}
	\caption{Model of the BBBC system.}
	\label{systemmodel}
	\vspace{-0.3cm}
\end{figure}

The BBBC system considered in this paper is shown in Fig.~\ref{systemmodel} and consists of a single-antenna carrier emitter (CE) as the RF source, a BD equipped with $M$ antennas, and a single-antenna reader. The CE emits a continuous sinusoidal carrier signal towards the BD, which includes backscatter antennas, information modulators, an energy harvester, a controller, and various other modules.
By periodically switching the load impedance connected to each antenna, the BD modifies the reflection coefficients, thereby modulating its information onto the incident carrier signal. As a result, the BD first captures a single-tone signal, which is then expanded into a broadband signal. Finally, the reader receives the backscattered signal and recovers the transmitted information of the BD. 
% Through the backscatter antenna, the BD can transmit its information to the reader by backscattering a portion of the incident signal, while harvesting energy from the remaining portion to empower the other modules.

The block fading channel model is adopted in this paper, wherein the channels remain constant over several blocks of transmitted symbols. Note that the single-tone signal is transmitted through the direct link from the CE to the reader and the forward link from the CE to the BD, while the broadband signal is transmitted through the backward link from the BD to the reader. Hence, we assume that both the direct and forward links are characterized by flat fading, whereas the backward link experiences frequency-selective fading with multipath spread. Denote by $h_\rmd$, $f_m$, and $\bg_m = [g_m(0), \ldots, g_m(l), \ldots, g_m(L_g - 1)]^\rmT$ the baseband channel impulse responses (CIRs) of the direct link, the forward link of the $m$-th BD antenna, and the backward link of the $m$-th BD antenna, respectively, where $L_g$ represents the number of channel taps in the backward link\footnote{Due to the close deployment of the BD antennas, we assume that the number of channel taps is identical for all antennas.}. Moreover, the stacked CIR from the CE to the BD is denoted by $\boldf = [f_1, \ldots, f_m, \ldots, f_M]^\rmT$.

\begin{figure}[t!]
	\centering    
	\includegraphics[width=0.99\columnwidth]{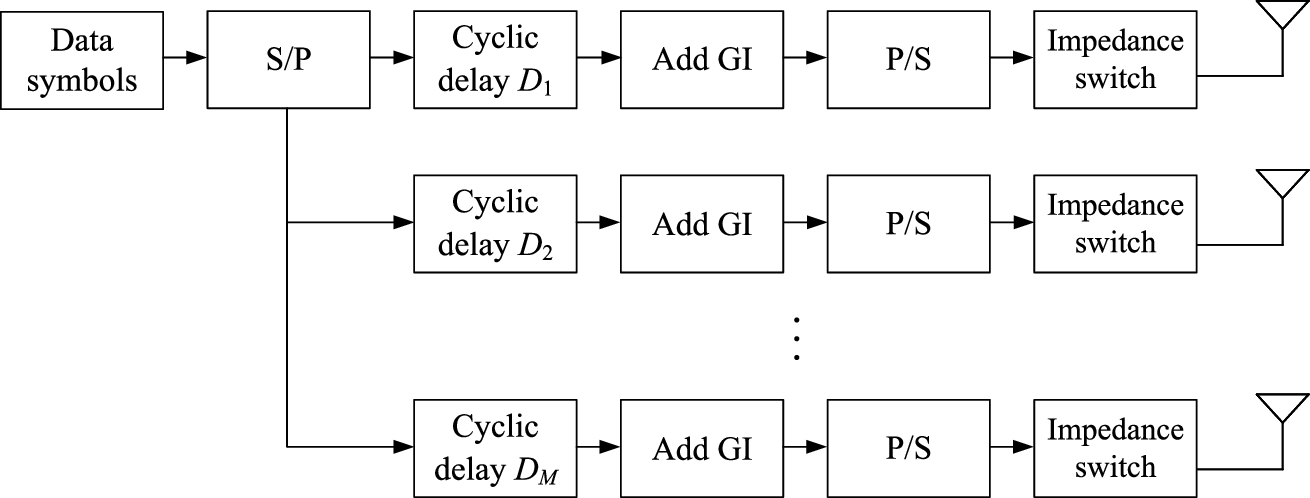}
	\caption{Block diagram of the information modulator at the BD.}
	\label{diagram}
	\vspace{-0.3cm}
\end{figure}

First, the CE emits a sinusoidal signal to the BD with transmit power of $p_\rmt$.
Thus, the signal captured by the $m$-th BD antenna is the attenuated sinusoidal signal, the baseband representation of which is given by $\sqrt{p_\rmt}f_m$.
To combat frequency-selective fading in the backward link and guarantee the transmission reliability of the BBBC system, the BD employs single-carrier block transmission and CDD. The diagram of the information modulator is illustrated in Fig.~\ref{diagram}. Specifically, the transmitted symbol stream is first divided into blocks. Let $N$ denote the size of each symbol block. After serial-to-parallel (S/P) conversion, the $k$-th symbol block can be written by 
\begin{align}
	\bc(k) = [c(kN), \ldots, c(kN + n), \ldots, c(kN + N - 1)]^\rmT,
\end{align}
where $c(n) \in \calA_c$ represents the $n$-th transmitted symbol, and $\calA_c$ is the modulation alphabet of the BD. Due to the passive backscattering nature of the BD, the amplitude of each transmitted symbol is limited by one\footnote{Some BDs are equipped with low-power reflection amplifiers, allowing symbol amplitudes greater than one but still constrained by the capacity of the reflection amplifier.}, i.e., $|c(n)| \le 1$. Here, we assume that the BD employs phase-shift keying (PSK) for modulation. The symbol block $\bc(k)$ is then conveyed to the branch connected to each antenna. Within each branch, a cyclic delay specific to each antenna is applied to $\bc(k)$. Specifically, the cyclically delayed signal for the $m$-th antenna is given by 
\begin{align}
	\bc_m(k) = \bT_m \bc(k),
\end{align}
where $\bT_m = \bT^{D_m}$, the $m$-th cyclic delay is denoted by $D_m$, and the cyclic delay matrix is given by 
\begin{align}
	\bT = \begin{bmatrix}
		\bm{0}_{N - 1}^\rmT & 1 \\
		\bI_{N - 1} & \mathbf{0}_{N - 1}
	\end{bmatrix}.
\end{align} 
Subsequently, a guard interval (GI) is inserted to prevent inter-block interference (IBI). There are several options for structuring the GI, including CP, ZP, and known training sequence (TS). For the low-power BD, ZP can provide two advantages. Firstly, it eliminates the need for continuous switching of load impedances, thereby reducing the energy consumption of the BD. Secondly, it can enhance energy harvesting during the ZP period by allowing the BD to select a reflection coefficient that maximizes energy capture from the incident signal. Hence, we adopt the ZP as the preferred GI\footnote{The results presented in this paper can be easily extended to the CP and TS GIs.}. The length of the ZP is denoted by $N_{\rm zp}$. Moreover, let $N_c = N + N_{\rm zp}$ and $\bT_{\rm zp} = [\bI_N, \bm{0}_{N \times N_{\rm zp}}]^\rmT$ denote the length of each symbol block after ZP insertion and the ZP insertion matrix, respectively. The symbol block with ZP insertion is given by
\begin{align} \label{zeropadding}
	\bc_m^{\rm zp}(k) & = \bT_{\rm zp} \bc_m(k) = \bT_{\rm zp} \bT_m \bc(k).
\end{align} 
The BD simultaneously switches the load impedance connected to each backscatter antenna to transmit symbols. The backscattered signal of the $m$-th BD antenna is obtained as $\sqrt{p_\rmt} f_m \bc_m^{\rm zp}(k)$.

Through the backward link, the cyclically shifted signals are superimposed at the reader. We assume that the ZP length is larger than the number of channel taps in the backward link, i.e., $N_{\rm zp} > L_g$. Denote by $\bh_m = [h_m(0), \ldots, h_m(l), \ldots, h_m(L_g - 1)]^\rmT$ the cascaded backscattered CIR via the $m$-th BD antenna, where $h_m(l) = f_m g_m(l)$. The baseband received signal can be expressed as
\begin{align} \label{receivedsignal}
	\bz(k) = \sqrt{p_\rmt}\left(h_\rmd \bm{1}_{N_c} + \sum_{m = 1}^{M}\bH_{1, m} \bc_m^{\rm zp}(k) \right) + \bu(k),
\end{align}
%\begin{align} \label{receivedsignal}
%	\bar{\bz}_k = \sqrt{p_\rmt}\left(h_\rmd \bm{1}_{N_c \times 1} + \sum_{m = 1}^{M}\bar{\bH}_m \bT_m \bc_k \right) + \bar{\bu}(k),
%\end{align}
where $\bH_{1, m}$ is a lower Toeplitz matrix, whose first column is given by $[\bh_m^\rmT, \bm{0}_{N_c - L_g}^\rmT]^\rmT$, and $\bu(k) \sim \calC\calN(\bm{0}_{N_c}, \sigma^2 \bI_{N_c})$ is additive white Gaussian noise (AWGN). In (\ref{receivedsignal}), the first term in the parenthesis is the interference from the direct link, which needs to be suppressed. The second term is the backscattered signal that contains the symbols transmitted by the BD.

\section{Receiver Design for BBBC} \label{receiverdesign}

\begin{figure}[t!]
	\centering    
	\includegraphics[width=0.99\columnwidth]{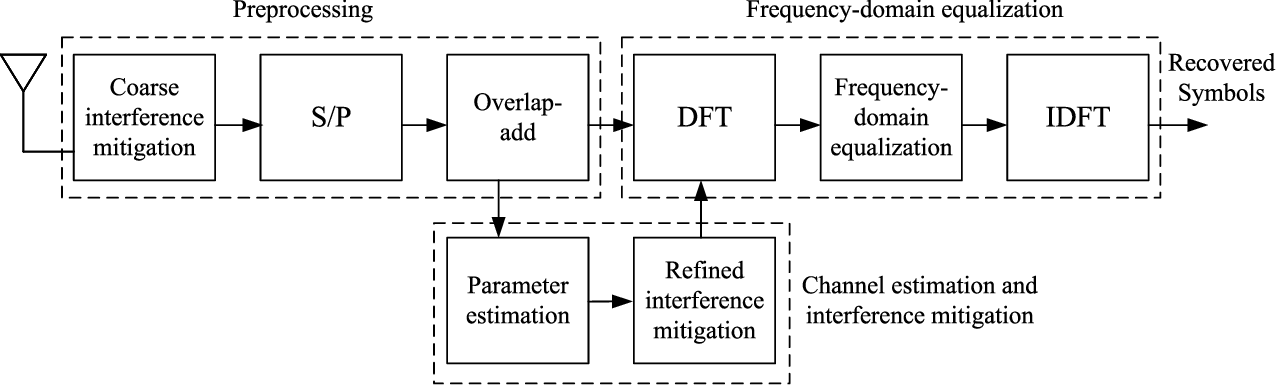}
	\caption{Block diagram of the receiver design at the reader.}
	\label{diagram_receiver}
	\vspace{-0.3cm}
\end{figure}

In this section, we propose a receiver design to recover the information transmitted by the BD, including preprocessing of the time-domain received signal, pilot-based parameter estimation, and frequency-domain equalization, as illustrated in Fig.~\ref{diagram_receiver}. 

\subsection{Preprocessing of Received Signal}
To narrow the gap in magnitudes between the direct-link interference and the backscattered signal, thereby ensuring the detection of the backscattered signal after the analog-to-digital converter (ADC), suppression of the direct-link interference in the analog domain is necessary\cite{VTCADC, MIMOBBC}. 
By averaging the received signal in the analog domain, the direct-link interference $\sqrt{p_\rmt} h_\rmd$ can be coarsely estimated and then removed from the received symbol block $\mathbf{z}(k)$.
% Therefore, by estimating and removing the mean value of the received signal, the direct-link interference can be coarsely suppressed before the ADC. 
Let $\bar{z}$ denote the mean value of the received signal, which is considered as an estimate of the direct-link interference. Following the coarse interference estimation and suppression, the baseband representation of the output signal is given by 
\begin{align}
    \hatbz(k) = \bz(k) - \bar{z} \bm{1}_{N_c}.
\end{align}
In practice, the duration of the received signal is limited, and thus, the estimate $\bar{z}$ contains a non-negligible error. As a result, the output signal $\hatbz(k)$ still includes residual interference, which remains to be finely mitigated in the digital domain.

\begin{figure}[t!]
    \centering    \includegraphics[width=0.99\columnwidth]{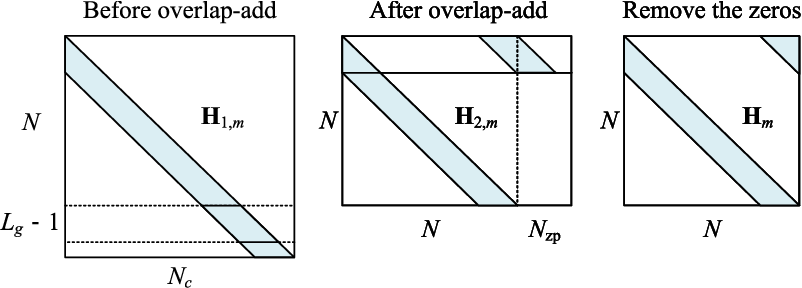}
	\caption{Channel matrix before and after overlap-add.}
	\label{overlapadd}
	\vspace{-0.3cm}
\end{figure}

The advantageous feature of circulant matrices, which allows them to be diagonalized via the discrete Fourier transform (DFT), facilitates the application of efficient frequency-domain equalization techniques to retrieve the information transmitted by the BD. To convert the channel matrix $\bH_{1, m}$ in (\ref{receivedsignal}) into a circular form, the overlap-add (OLA) method can be employed \cite{CPorZP}, as illustrated in Fig.~{\ref{overlapadd}}. Specifically, we assume that the number of channel taps $L_g$ is a priori known to the reader via channel sounding. After time synchronization, the reader divides the received signal into blocks. In each block, the segment of the received GI that experiences interference from the transmitted symbols, represented by $\hatbz_2(k) \in \bbC^{(L_g - 1) \times 1}$, is first padded with $(N - L_g + 1)$ zeros. Then, the padded segment is combined with the upper $N \times 1$ segment of $\hatbz(k)$, represented by $\hatbz_1(k)$. Hence, the output signal can be expressed as
% Specifically, the received signal $\hatbz(k)$ is segmented into two parts: the upper $N \times 1$ segment $\hatbz_1(k)$, and the lower $N_{\rm zp} \times 1$ segment $\hatbz_2(k)$. Then, the lower segment $\hatbz_2(k)$ is padded with $N - N_{\rm zp}$ zeros, and combined with the upper segment $\hatbz_1(k)$, resulting in
\begin{align} \label{receivedsignal2}
	\by(k) & = \hatbz_1(k) + \begin{bmatrix}
		\hatbz_2(k) \\
		\bm{0}_{N - L_g + 1} 
	\end{bmatrix} \nonumber \\
	& = \sqrt{p_\rmt} \left(\sum_{m = 1}^{M} \bH_{2, m} \bc_m^{\rm zp}(k) + \Delta h_\rmd \balpha_N \right) + \bu_{\rm eq}(k),
\end{align}
%\begin{align} \label{receivedsignal2}
%	\by(k) & = \hatbz_1(k) + \begin{bmatrix}
%		\hatbz_2(k) \\
%		\bm{0}_{N - N_{\rm zp}} 
%	\end{bmatrix}  \nonumber \\
%	& = \sqrt{p_\rmt} \left(\sum_{m = 1}^{M} \bH_{2, m} \bT_m \bc(k) + \Delta \bh_\rmd \right) + \bu_{\rm eq}(k) \nonumber \\
%	& = \sqrt{p_\rmt} \left(\bH \bc(k) + \Delta \bh_\rmd\right) + \bu_{\rm eq}(k),
%\end{align}
where the channel matrix is given by $\bH_{2, m} = \bH_{1, m}(1 : N, :) + [(\bH_{1, m} (N + 1 : N + L_g - 1, :))^\rmT, \bm{0}_{N_c \times (N - L_g + 1)}]^\rmT$, $\sqrt{p_\rmt}\Delta h_\rmd \balpha_N$ represents the residual interference after OLA with $\Delta h_\rmd = h_\rmd - \frac{\bar{z}}{\sqrt{p_\rmt}}$ and $\balpha_N = [2\cdot \bm{1}_{N_{\rm zp}}^\rmT, \bm{1}_{N - L_g + 1}^\rmT]^\rmT$, and $\bu_{\rm eq}(k) \sim \calC\calN(\bm{0}_N, \diag(\sigma^2 \balpha_N))$ is the equivalent noise. With the substitution of (\ref{zeropadding}) into (\ref{receivedsignal2}), the received signal is obtained as
\begin{align}\label{receivedsignal3}
	\by(k) & = \sqrt{p_\rmt} \left(\sum_{m = 1}^{M} \bH_m \bT_m \bc(k) + \Delta h_\rmd \balpha_N \right) + \bu_{\rm eq}(k) \nonumber \\
	& = \sqrt{p_\rmt} \left(\bH_{\rm eq} \bc(k) + \Delta h_\rmd \balpha_N \right) + \bu_{\rm eq}(k),
\end{align}
where $\bH_m = \bH_{2, m}\bT_{\rm zp}$ and $\bH_{\rm eq} = \sum_{m = 1}^{M} \bH_m \bT_m$ are both circulant matrices with their first columns given by $[\bh_m^\rmT, \bm{0}_{N - L_g}^\rmT]^\rmT$ and 
\begin{align} \label{equivalentCIR}
	\bh_{\rm eq} = \sum_{m = 1}^{M} \bT_m \begin{bmatrix}
		\bh_m \\
		\bm{0}_{N - L_g}
	\end{bmatrix},
\end{align}
respectively. As a result, the convolution between each BD symbol block and the cascaded backscattered-link channel becomes a circular convolution thanks to the OLA operation.

\begin{remark}
    The equivalent CIR in (\ref{equivalentCIR}) is the combination of the delayed CIRs of all the BD antennas. To enhance the diversity gain of the system, the equivalent CIR should have as many independent taps as possible. This indicates that the cyclic delay of each BD antenna can be designed to ensure that the combination of these delayed CIRs results in minimal overlap to maximize the number of independent taps of the equivalent channel. 
    % , which means that each tap in (\ref{equivalentCIR}) should avoid being the summation of channel taps related to different BD antennas. 
    Hence, the cyclic delay can be chosen as $D_m = (m - 1) D$, where the cyclic delay between adjacent antennas, $D$, needs to satisfy $L_g \le D \le \lfloor \frac{N}{M} \rfloor$. The resulted equivalent CIR can be written as
    \begin{multline}
        \bh_{\rm eq} = \left[\bh_1^\rmT, \bm{0}_{D - L_g}^\rmT, \ldots, \bh_m^\rmT, \bm{0}_{D - L_g}^\rmT, \ldots, \right. \\
        \left.  \bh_M^\rmT, \bm{0}_{N - (M - 1) D} \right]^\rmT 
        = \bM_1 \bh,
    \end{multline}
    where $\bM_1 = [\bI_{M L_g}(:, 1 : L_g), \bm{0}_{M L_g \times (D - L_g)}, \ldots, \bI_{M L_g}(:, (m - 1)L_g + 1 : m L_g), \bm{0}_{M L_g \times (D - L_g)}, \ldots, \bI_{M L_g}(:, (M - 1)L_g : M L_g), \bm{0}_{M L_g \times (N - (M - 1)D)}]^\rmT \in \bbC^{N\times ML_g}$ and $\bh = \left[\bh_1^\rmT, \ldots, \bh_m^\rmT, \ldots, \bh_M^\rmT\right]^\rmT$. Moreover, this also requires the block size $N$ to be no less than $M L_g$, i.e., $N \ge M L_g$.
\end{remark}

\subsection{Pilot-Based Channel Estimation and Refined Interference Suppression}

Before transmitting the data payload, the BD sends pilot symbols, known to both the BD and the reader, to help the reader acquire the equivalent CIR and suppress the residual interference. The pilot symbols are similarly modulated and conveyed as the data symbols. Let $N_\rmp$ denote the number of pilot symbol blocks, and let $\bc_\rmp(k_\rmp) \in \bbC^{N\times 1}$ denote the $k_\rmp$-th pilot symbol block. Correspondingly, the received pilot symbol blocks can be collectively represented as
\begin{align}
	\by_{\rm tr} \!=\! \underbrace{\sqrt{p_\rmt} \! \begin{bmatrix}
		\bC_\rmp(1) \!&\! \balpha_N \\
		\vdots \!&\! \vdots \\
		\bC_\rmp(N_\rmp) \!&\! \balpha_N
	\end{bmatrix} \begin{bmatrix}
	    \bM_1 & \bm{0}_{N} \\
        \bm{0}_{ML_g}^\rmT & 1
	\end{bmatrix}}_{\bM_2} \! \! \begin{bmatrix}
	\bh_{\rm eq} \\
	\Delta h_\rmd
	\end{bmatrix} \!+\! \bu_{\rm tr},
\end{align}
where $\by_{\rm tr} = [\by^\rmT(1), \ldots, \by^\rmT(N_\rmp)]^\rmT$ is the stacked received pilot symbol blocks, $\bu_{\rm tr} = [\bu_{\rm eq}^\rmT(1), \ldots, \bu_{\rm eq}^\rmT(N_\rmp)]^\rmT$ is the noise, and $\bC_\rmp(k_\rmp)$ is a circulant matrix with its first column given by $\bc_\rmp(k)$. To accurately estimate $\bh_{\rm eq}$ and $\Delta h_\rmd$, the number of pilot symbol blocks, $N_\rmp$, needs to be larger than 1. Considering the limited power consumption at the BD, we set $N_\rmp = 2$. Thus, the estimation can be performed using a least squares (LS) estimator given by
\begin{align} \label{LSestimation}
	\begin{bmatrix}
		\bh_{\rm eq}^{\rm est} \\
		\Delta h_\rmd^{\rm est}
	\end{bmatrix} = \left(\bM_2^\rmH \bM_2\right)^{-1} \bM_2^\rmH \by_{\rm tr},
\end{align}
where $\bh_{\rm eq}^{\rm est}$ and $\Delta h_\rmd^{\rm est}$ are the estimates of $\bh_{\rm eq}$ and $\Delta h_\rmd$, respectively.
Assuming that the block size is much larger than the number of taps for the backward link, i.e., $N \gg L_g$, the covariance of the estimate in (\ref{LSestimation}) can be approximated as
\begin{align}
	& \bSigma_{\rm est} \approx \frac{N_c \sigma^2}{N} \left(\bM_2^\rmH\bM_2\right)^{-1} \nonumber \\
	& = \frac{\sigma^2}{p_\rmt} \begin{bmatrix}
		\sum_{k = 0}^{1} \! \bC_\rmp^\rmH(k)\bC_\rmp(k) \!&\! \sum_{k = 0}^{1} \! \bC_\rmp^\rmH(k) \balpha_N \\
		\balpha_N^\rmH \sum_{k = 0}^{1}\bC(k) \!&\! 2 \|\balpha_N\|^2
	\end{bmatrix}^{-1}\!\!.
\end{align}
To minimize the mean square error (MSE) of the estimates, the non-diagonal elements of $\bSigma_{\rm est}$ should be zeros, implying that the first pilot symbol block should be the negative version of the second pilot symbol block. Thus, we set $\bC_\rmp(0) = - \bC_\rmp(1) = \bC_\rmp$. The MSEs of the estimates for $\bh$ and $\Delta h_\rmd$ are obtained as
\begin{align}
	& \epsilon_h = \bbE[\|\bh_{\rm eq} - \bh_{\rm eq}^{\rm est}\| ^ 2] = \frac{\sigma^2}{2 p_\rmt} \tr \left(\left(\bC_\rmp^\rmH \bC_\rmp\right)^{-1}\right), \\
	& \epsilon_\rmd = \bbE[\|\Delta h_\rmd - \Delta h_\rmd^{\rm est}\|^2 ] = \frac{\sigma^2}{2p_\rmt (N + 3(L_g - 1)))}. 
\end{align}
Note that the MSE of the estimate of $\bh_{\rm eq}$ is influenced by the pilot symbol block $\bC_\rmp$. To minimize $\epsilon_h$, the pilot symbol block needs to satisfy $\bC_\rmp^\rmH \bC_\rmp = \bI_N$ \cite{WEN}. Zadoff-Chu sequences \cite{ZadoffChu}, known for their good auto-correlation properties, are ideal candidates for the pilot symbol block. However, given the inherent constraints of the low-cost BD, which is restricted to switching reflection coefficients from a specific finite set for backscattering, the implementation of Zadoff-Chu sequences poses challenges. Thus, we quantize the Zadoff-Chu sequence according to the modulation alphabet $\calA_c$ and use it as the pilot symbol block.

As a result, given the estimates of $\bh_{\rm eq}^{\rm est}$ and $\Delta h_\rmd^{\rm est}$, the reader can effectively remove the residual interference and further facilitate frequency-domain equalization.
We assume that the estimates are perfect for the following receiver design.

%\begin{remark}
%	Due to the constraints of cost and power consumption, the BD cannot be equipped with phase shifters like conventional active transmitters to generate symbols with arbitrary phase shifts. On the contrary, the modulation order of the BD is determined by the limited load impedances connected to each BD antenna. Therefore, to enable the BD to transmit the aforementioned pilot signal, the block size $N$ is required to be a multiple of the modulation order of the BD.
%\end{remark}

\subsection{Frequency-Domain Equalization}

Recall that channel matrix $\bH_{\rm eq}$ in (\ref{receivedsignal3}) can be decomposed as $\bH_{\rm eq} = \bW^\rmH \bLambda \bW$, where $\bW$ is the unitary DFT matrix with its $(p, q)$-th element given by $\frac{1}{\sqrt{N}}\exp(-j2\pi(p - 1)(q - 1) / N)$.
Hence, to facilitate frequency-domain equalization at the reader, the received signal in (\ref{receivedsignal3}) is transformed to the frequency domain via DFT, which yields
\begin{align}\label{receivedsignal4}
    \tilby (k) & = \frac{1}{\sqrt{p_\rmt}}\bW \left( \by(k) - \sqrt{p_\rmt} \Delta h_\rmd \balpha_N \right) \nonumber \\
    & = \mathbf{\Lambda}\bW \bc(k) + \tilbu_{\rm eq}(k),
\end{align}
where $\tilbu_{\rm eq}(k) = \bW \bu_{\rm eq}(k) / \sqrt{p_\rmt}$ is the frequency-domain noise. 
We observe that each element of $\bW \bc(k)$ experiences a single-tap channel, which is given by $\lambda_n = \sum_{l = 0}^{N - 1} h(l) \exp(-j2\pi n(l - 1) / N)$ for the $n$-th DFT grid.
Therefore, a linear equalizer, denoted by $\bG$, can be employed to recover $\bc(k)$ based on the received signal model in (\ref{receivedsignal4}). Specifically, the zero-forcing (ZF) and minimum-mean-square-error (MMSE) equalizers can be expressed respectively as 
\begin{gather}
    \bG_{\rm ZF} =  \bW^\rmH \bLambda^{-1}, \\
    \bG_{\rm MMSE} =  \bW^\rmH \bLambda^\rmH\left(\bLambda^\rmH \bLambda +  \bar{\gamma}^{-1} \bI\right)^{-1},
\end{gather}
where $\bar{\gamma} = p_\rmt \Delta\gamma / \sigma^2$ and $\Delta \gamma = N / (N + L_g - 1))$ represents the signal-to-noise ratio (SNR) loss since the elements of $\tilbu_{\rm eq}(k)$ follow distribution $\calC\calN(0, 1/\bar{\gamma})$. Hence, the output of the linear frequency-domain equalizer can be written as
\begin{align}
	\hatbc(k) = \bG \tilby(k).
\end{align}
We assume $\bbE[\bc(k)\bc^\rmH(k)] = \bI_N$. Meanwhile, under the assumption that $N \gg L_g$, the elements of $\tilbu_{\rm eq} (k)$ are independent of each other. Hence, the output SNR of the linear frequency-domain equalizers can be derived as
\begin{gather}
    \gamma_{\rm ZF} = \frac{\bar{\gamma}N}{\sum_{n = 1}^{N} \left|\lambda_n\right|^{-2}}, \label{snrzf}\\
    \gamma_{\rm MMSE} = \left(\frac{1}{N}\sum_{n = 1}^{N}\frac{1}{1 + \bar{\gamma} \left|\lambda_n \right|^2}\right)^{-1} - 1.\label{snrmmse}
\end{gather}
Note that the output SNR is identical for each symbol within a single transmission block. Furthermore, when employing QPSK modulation at the BD, the bit error rate (BER) performance can be calculated as $P_{\rm err} = \calQ(\sqrt{\gamma_{\rm ZF/MMSE}})$, where the Q-function is given by $\calQ(x) = \frac{1}{\sqrt{2\pi}} \int_{x}^{\infty} \exp(-u^2 / 2)\rmd u$.

Compared to the above linear equalizers, block-iterative generalized decision-feedback equalizion (BI-GDFE) can achieve enhanced performance by effective ISI cancellation \cite{BIGDFE}. Specifically, in each iteration, the symbol decisions from the previous iteration are used to regenerate the ISI, which is then removed from the received signal following equalization. Denote by $\mathbf{K}_l$ and $\mathbf{D}_l$ the feedforward equalizer (FFE) and the feedback equalizer (FBE) in the $l$-th iteration, respectively. Then, the estimate of $\mathbf{c}(k)$ in the $l$-th iteration is given by
\begin{align} \label{bigdfe}
    \widehat{\mathbf{c}}^{\{l\}}(k) = \mathbf{K}_l^\mathrm{H} \tilby(k) + \mathbf{D}_l \widehat{\mathbf{c}}^{\{l - 1\}}(k),
\end{align}
where $\widehat{\mathbf{c}}^{\{l\}}(k)$ is the estimate of $\mathbf{c}(k)$ in the $l$-th iteration. In (\ref{bigdfe}), the first term represents the equalized signal, while the second term represents the removal of the residual ISI. From \cite{BIGDFE}, the FFE and FBE are respectively given by
\begin{align}
    \mathbf{K}_l & = \mathbf{F}_l \mathbf{W}, \\
    \mathbf{B}_l & = \rho_{l - 1}\mathbf{W}^\mathrm{H} \mathbf{B}_l \mathbf{W},
\end{align}
where $\rho_{l - 1}$ is the input-decision correlation (IDC) coefficient between the transmitted symbol block $\mathbf{c}(k)$ and the hard decisions derived based on $\widehat{\mathbf{c}}^{\{l - 1\}}(k)$, the basic FFE and FBE are respectively given by 
\begin{align}
    \mathbf{F}_l & = \left((1 - \rho_{l - 1}^2)\mathbf{\Lambda}^\mathrm{H}\mathbf{\Lambda} + \bar{\gamma}^{-1} \mathbf{I}_N\right)^{-1} \mathbf{\Lambda}, \\
    \mathbf{B}_l & = \alpha_l \mathbf{I}_N - \mathbf{F}_l^\mathrm{H} \mathbf{\Lambda},
\end{align}
and $\alpha_l = \tr(\mathbf{F}_l^\mathrm{H} \mathbf{\Lambda}) / N$. As the iterations progress, the accuracy of the decisions based on the estimate in (\ref{bigdfe}) increases, thereby the IDC $\rho_l$ approaches unity, which leads to a reduction in the residual ISI. As a result, the performance of the BI-GDFE is enhanced iteratively, resulting in increasingly reliable signal detection of $\mathbf{c}(k)$.

\section{Diversity Analysis of BBBC} \label{diversityanalysis}

Non-linear equalizers, such as BI-GDFE, have been shown to have the potential to approach the matched filter bound (MFB), which serves as a performance upper bound \cite{BIGDFE}. To analyze the achievable diversity order of our proposed BBBC system, the following analysis will focus on the MFB. Specifically, we assume perfect cancellation of ISI and a coherent combination of the desired signals to derive the MFB. Thus, for the MFB, the output SNR can be expressed as
\begin{align} \label{snr}
	\gamma_c = \frac{ p_\rmt \Delta \gamma \|\bh\|^2}{\sigma^2} = \frac{ p_\rmt \Delta \gamma \sum_{m = 1}^{M}\left|f_m\right| ^ 2 \left\|\bg_m\right\|^2}{\sigma^2}. 
\end{align}
From (\ref{snr}), it can be observed that deploying more antennas at the BD can lead to a higher SNR. This enhancement is due to the passive nature of the BD, which merely backscatters the signal incident from the CE rather than generating a power-limited RF signal. Additional antennas enhance the capability of the BD to capture and backscatter the incident signal.

Moreover, the diversity order characterizes the decay of the average BER for increasing transmit power $p_\rmt$. Since the direct computation of the average BER is not tractable, the diversity order is analyzed in terms of the outage probability, which exhibits identical diversity gain results as the average BER. Let $P_{\rm out}(p_\rmt)$ denote the outage probability. The diversity order is defined as  
\begin{equation} \label{definition}
	d = - \lim_{p_\rmt \to \infty} \frac{\log_{10}P_{\rm out} (p_\rmt)}{p_\rmt}.
\end{equation}

In the following, we will consider two typical scenarios for the BBBC system: the general channel conditions where both the forward and backward links experience Rayleigh fading channels and the special channel conditions where the forward link experiences a LoS channel while the backward link remains Rayleigh fading. We first present the channel models for these scenarios, followed by the analysis of the outage probability and diversity order.

\subsection{General Channel Conditions}
We consider the scenario where the forward- and backward-link channels follow the Rayleigh fading channel model with an uniform power profile, i.e., $f_m \sim \calC\calN(0, \beta_1)$ for the forward link and $g_m(l) \sim \calC\calN(0, \beta_2 / L_g)$ for the backward link, where $\beta_1$ and $\beta_2$ are the corresponding large-scale fading coefficients, respectively. We first analyze the case with a single antenna ($M = 1$), which is extended to the case where $M > 1$\footnote{While \cite{GainsMBC} provides a comprehensive investigation of the two-way pinhole channel assuming flat fading for both the forward and backward links, our study extends this analysis by considering scenarios with frequency-selective fading, thus establishing new channel models for backscatter communications.}.
\subsubsection{Single-Antenna Case}
Based on the above assumptions, $|f|^2$ follows an exponential distribution characterized by parameter $\beta_1$, while $\|\bg\|^2$ follows a scaled Chi-squared distribution with $2 L_g$ degrees of freedom. The probability density functions (PDFs) of these distributions are given by
\begin{equation}
	f_{\left|f\right|^2}(x) = \frac{1}{\beta_1}\exp\left(-\frac{x}{\beta_1}\right),
\end{equation}
\begin{equation}
	f_{\left\|\bg\right\|^2}(y) \!=\! \frac{1}{2^{L_g}\Gamma(L_g)} \left(\frac{2L_g}{\beta_2}\right)^{L_g}y^{L_g - 1} \exp\left(-\frac{L_g}{\beta_2}y\right),
\end{equation}
where $\Gamma(L_g)$ denotes the Gamma function. Furthermore, the PDF of the product $|f|^2 \|\bg\|^2$ can be derived as
\begin{align} \label{pdfexpression}
	& f_{\left|f\right|^2\left\|\bg\right\|^2}(z) = \int_{0}^{\infty}\frac{1}{u}f_{\left|f\right|^2}\left(\frac{z}{u}\right)f_{\left\|\bg\right\|^2}(u) \rmd u \nonumber \\
	& = \frac{1}{\Gamma(L_g)\beta_1}\left(\frac{L_g}{\beta_2}\right)^{L_g}\int_{0}^{\infty} u^{L_g - 2} \exp\left(-\frac{z}{\beta_1 u} -\frac{L_g}{\beta_2}u\right) \rmd u \nonumber \\
	& \overset{(a)}{=} \frac{L_g}{\Gamma(L_g)\beta_1\beta_2}\int_{0}^{\infty} t^{L_g - 2} \exp\left(- t - \frac{L_g z}{\beta_1\beta_2 t} \right) \rmd t \nonumber \\
	& \overset{(b)}{=} \frac{2}{\Gamma(L_g)} \left(\frac{L_g}{\beta_1 \beta_2}\right)^{\frac{L_g \!+\! 1}{2}} z^{\frac{L_g \!-\! 1}{2}} K_{L_g \!-\! 1}\left(2 \left(\frac{L_g z}{\beta_1 \beta_2}\right)^{\frac{1}{2}} \right),
\end{align}
where $(a)$ results from the substitution of $u = \beta_2 t / L_g$, and $(b)$ is due to the fact that $\int_{0}^{\infty}t^{\nu - 1} \exp\left(-t - \frac{\mu^2}{4t}\right) \rmd t = 2 \left(\frac{\mu}{2}\right)^\nu K_{-\nu}(\mu)$ with $K_{-\nu}(\mu) = K_\nu(\mu)$ representing the modified Bessel function of second kind with order $\nu$. From the expression in (\ref{pdfexpression}), we notice that the equivalent channel gain $|f|^2 \|\bg\|^2$ follows the generalized-$K$ distribution \cite{GK}, which is characterized by the product of two independent gamma distributions. The general PDF and cumulative density function (CDF) of the generalized-$K$ distribution are given by
\begin{multline}
\label{kgpdf}
	f_x (x; k, m, \Omega) = \frac{2}{\Gamma(k) \Gamma(m)} \left(\frac{km}{\Omega}\right)^{\frac{k + m}{2}} \\
    x^{\frac{k + m}{2} - 1} K_{k - m}\left(2\left(\frac{km}{\Omega}x\right)^{\frac{1}{2}}\right),
\end{multline}
\begin{equation}
	F_x (x; k, m, \Omega) = \frac{1}{\Gamma(k) \Gamma(m)} G_{1, 3}^{2, 1}\left(\frac{k m x}{\Omega} \; \middle| \begin{array}{c}
		1 \\ [-1ex]
		k, m, 0
	\end{array}\right),
\end{equation}
where $k \ge 0$ and $m \ge 0$ are the shape parameters, $\Omega = \bbE[x]$ is the mean of $x$, and $G(\cdot)$ represents Meijer's G function.
For the PDF specified in (\ref{pdfexpression}), these parameters are given by $k_1 = L_g$, $m_1 = 1$, $\Omega_1 = \beta_1 \beta_2$. The variance of the equivalent channel gain is derived as
\begin{align}
    \bbV[|f|^2 \|\bg\|^2] = \Omega_1^2 \left(\frac{k_1 \!+\! m_1 \!+\! 1}{k_1 m_1}\right) = \beta_1^2 \beta_2^2 \left(1 \!+\! \frac{2}{L_g}\right).
\end{align}
Note that the variance decreases as the number of channel taps $L_g$ increases.
The target SNR required to detect the symbol block $\bc(k)$ is denoted by $\gamma_{\rm th} = 2 ^ R - 1$, where $R$ bps/Hz is the target rate for the BD. The outage probability is obtained as
\begin{align} \label{op1}
	P_{\rm out} & = \Pr\left\{|f|^2 \|\bg\|^2 \le \frac{\sigma^2 \gamma_{\rm th}}{p_\rmt \Delta \gamma}\right\} \nonumber \\
	& = \frac{1}{\Gamma(L_g)} G_{1, 3}^{2, 1}\left(\frac{\sigma^2 \gamma_{\rm th} L_g }{p_\rmt \Delta \gamma} \; \middle| \begin{array}{c}
		1 \\ [-1ex]
		L_g, 1, 0
	\end{array}\right).
\end{align}
Given the expression for the outage probability, we obtain the following lemma.
\begin{mylem} \label{pro1}
	Considering the generalized-$K$ fading channel characterized by its parameters $k$, $m$, and $\Omega$, the diversity order of the BBBC system is equal to parameter $m$.
\end{mylem}
\begin{IEEEproof}
Using the binomial series expansion of the moment generating function (MGF) of the received SNR, \textit{Lamma} 1 can be proved using the approach proposed in \cite{Chintha}. Here, we provide an alternative proof in terms of the outage probability. The expression for the outage probability, obtained directly from the CDF involving Meijer's G function, raises challenges for further analysis. Therefore, an alternative formulation of the outage probability is given by
	\begin{align}
		P_{\rm out} = \int_{0}^{a_1 p_\rmt^{-1}}  f_z(z; k, m, \Omega) \rmd z,
	\end{align}
where $a_1 = \sigma^2 \gamma_{\rm th} / \Delta \gamma$. As the transmit power $p_\rmt$ tends to infinity, the outage probability approaches zero due to $\lim_{p_\rmt \to \infty} a_1 p_\rmt^{-1} = 0$. Hence, according to the definition in (\ref{definition}), the diversity order can be determined by 
\begin{align}
	d & \overset{(c)}{ = } -\lim_{p_\rmt \to \infty} \frac{p_\rmt}{P_{\rm out}(p_\rmt)}P_{\rm out}'(p_\rmt) \nonumber \\
	& = \lim_{p_\rmt \to \infty}\frac{\frac{2(a_1 a_2)^{\frac{k + m}{2}}}{\Gamma(k)\Gamma(m)} p_\rmt^{-\frac{k + m}{2}} K_{k - m}\left(2\left(a_1 a_2\right)^{\frac{1}{2}} p_\rmt^{-\frac{1}{2}}\right)}{\frac{2 a_2^{\frac{k + m}{2}}}{\Gamma(k)\Gamma(m)} \int_{0}^{a_1 p_\rmt^{-1}} z^{\frac{k + m}{2} - 1} K_{k - m}\left(2\left(a_2 z\right)^{\frac{1}{2}}\right) \rmd z} \nonumber \\
	& = a_1^{\frac{k + m}{2}} \lim_{p_\rmt \to \infty} \frac{p_\rmt^{-\frac{k + m}{2}} K_{k - m}\left(2\left(a_1 a_2\right)^{\frac{1}{2}} p_\rmt^{-\frac{1}{2}}\right)}{\int_{0}^{a_1 p_\rmt^{-1}} z^{\frac{k + m}{2} - 1} K_{k - m}\left(2\left(a_2 z\right)^{\frac{1}{2}}\right) \rmd z} \nonumber \\
	& \overset{(d)}{=} \lim_{p_\rmt \to \infty} \frac{\frac{\Gamma(k - m)}{2} a_1^{-m} a_2^{-\frac{k - m}{2}} p_\rmt^{-m}}{\int_{0}^{a_1 p_\rmt^{-1}} z^{\frac{k + m}{2} - 1} K_{k - m}\left(2\left(a_2 z\right)^{\frac{1}{2}}\right) \rmd z} \nonumber \\
	& \overset{(e)}{=} \lim_{p_\rmt \to \infty} \frac{\frac{\Gamma(k - m)}{2} (a_1 a_2)^{-\frac{k - m}{2}} m p_\rmt^{- m - 1}}{p_\rmt^{-\frac{k + m}{2} - 1} K_{k - m}\left(2 (a_1 a_2)^{\frac{1}{2}} p_\rmt^{-\frac{1}{2}}\right)} \nonumber \\ 
	& \overset{(f)}{=} m,
\end{align}
where the first-order derivative is given by
\begin{align}
	P_{\rm out}'(p_\rmt) \!=\! -\frac{2(a_1 a_2)^{\frac{k + m}{2}}}{\Gamma(k)\Gamma(m)} p_\rmt^{-\frac{k + m}{2} - 1} K_{k \!-\! m}\left(2 (a_1 a_2)^{\frac{1}{2}} p_\rmt^{-\frac{1}{2}}\right),
\end{align}
and $a_2 = k m / \Omega $. Here, $(c)$ and $(e)$ result from L'Hôpital's rule, while $(d)$ and $(f)$ are obtained from the asymptotic behavior of $K_\nu(\mu)$, which can be approximated $\frac{\Gamma(\nu)}{2}\left(\frac{2}{\mu}\right)^\nu$ as $\mu$ tends to zero. This concludes the proof.
\end{IEEEproof}
According to \emph{Lemma}~\ref{pro1}, the diversity order of the BBBC system is given by $1$ in the single-antenna case. Although multiple propagation paths are available for transmitting the BD signal over the backward link via each antenna, the channel fading in the forward link restricts the diversity order to 1 when using a single antenna, thereby preventing the acquisition of frequency diversity. To improve the diversity performance of the BBBC system, multiple antennas can be deployed at the BD, as outlined below.

\subsubsection{Multi-Antenna Case}
When $M > 1$, it is challenging to derive the exact PDF of the summation $\sum_{m = 1}^{M} \left|f_m\right| ^ 2 \left\|\bg_m\right\|^2$. According to \cite{sumGK}, the sum of generalized-$K$ distributions can be approximated by another generalized-$K$ distribution. Therefore, the PDF of the equivalent channel gain $\sum_{m = 1}^{M} \left|f_m\right| ^ 2 \left\|\bg_m\right\|^2$ has the same form as in (\ref{kgpdf}), and its parameters can be derived as
\begin{align} \label{parameters}
	k_\rmM = M L_g + \epsilon, \ m_\rmM = M, \ {\rm and} \ \Omega_\rmM = M \beta_1 \beta_2,
\end{align}
where $\epsilon$ is the adjustment parameter given by
\begin{align}
	\epsilon = (M - 1)\frac{-0.127-0.95k_1 -0.0058m_1}{1 + 0.00124k_1 + 0.98m_1}.
\end{align}
Thus, the outage probability can be obtained as 
\begin{align} \label{op_expressions_general_case}
	& P_{\rm out}  = \Pr\left\{\sum_{m = 1}^{M}|f_m|^2 \|\bg_m\|^2 \le a_1 p_\rmt^{-1}\right\} \nonumber \\
	& = \! \frac{1}{\Gamma(K_\rmM) \Gamma(m_\rmM)} G_{1, 3}^{2, 1} \! \left(\frac{k_\rmM m_\rmM a_1 p_\rmt^{-1}}{\Omega_\rmM} \; \!\middle|\! \begin{array}{c}
		1 \\ [-1ex]
		k_\rmM, m_\rmM, 0
	\end{array}\!\right).
\end{align}
According to \emph{Proposition}~\ref{pro1}, we conclude that the diversity order of the BBBC system increases to $M$ in the multi-antenna case. This diversity gain arises from each antenna creating a signal replica that is transmitted with a specific delay. Moreover, note that employing additional antennas at the BD not only leads to an increase in SNR when dealing with fixed channels but also introduces spatial diversity to improve the transmission reliability over fading channels.

\subsection{Special Channel Conditions}
In addition to the case of the general channel conditions analyzed above, we consider the special channel conditions, where the forward-link channel is a deterministic LoS channel, while the backward-link channel follows the Rayleigh fading channel model as mentioned ealier\footnote{In certain applications, the locations of the CE and the BD may be fixed to ensure a strong LoS component between them, thereby enabling reliable capture of the carrier signal at the BD.}. Accordingly, the forward-link channel is given by
\begin{equation}
	\boldf = \sqrt{\beta_1} \left[1, e^{-j2\pi r \sin\theta}, \ldots, e^{-j2\pi(M - 1) r \sin\theta}\right],
\end{equation}
where $\theta$ is the angle of arrival (AoA) and $r$ is the ratio of the antenna spacing to the wavelength. The SNR of the backscattered signal can be rewritten as
\begin{equation}
	\gamma_c = \frac{p_\rmt N \beta_1 \sum_{m = 1}^{M} \left\|\bg_m\right\|^2}{\sigma^2 (N + L_g - 1)},
\end{equation}
where $2 L_g \sum_{m = 1}^{M} \|\bg_m\|^2 / \beta_2$ follows a Chi-squared distribution with $2 M L_g$ degrees of freedom. Hence, the outage probability can be derived as
\begin{align} \label{op_expression_special_case}
	P_{\rm out}(p_\rmt) & = \Pr\left\{\frac{2 L_g \sum_{m = 1}^{M} \|\bg_m\|^2}{\beta_2} \le a_3 p_\rmt^{-1}\right\} \nonumber \\
	& = \int_{0}^{a_3 p_\rmt^{-1}} \frac{1}{2 ^{M L_g} \Gamma(M L_g)} z^{M L_g - 1} e^{-\frac{z}{2}} \rmd z \nonumber \\
	& = \frac{1}{\Gamma(M L_g)}\gamma\left(M L_g, \frac{a_3}{2 p_\rmt}\right),
\end{align}
where $a_3 = 2 a_1 L_g /(\beta_1 \beta_2)$ and $\gamma(x, y)$ is the lower incomplete Gamma function. We observe that the outage probability tends to zero with the transmit power $p_\rmt$ approaching infinity. Hence, the diversity order is given by 
\begin{align} \label{diversity2}
	d & = -\lim_{p_\rmt \to \infty} \frac{p_\rmt}{P_{\rm out}(p_\rmt)}P_{\rm out}'(p_\rmt) \nonumber \\
	& = \lim_{p_\rmt \to \infty} \frac{ a_3^{M L_g} p_\rmt ^{-M L_g} \exp\left(- \frac{1}{2} a_3 p_\rmt^{-1} \right)}{\int_{0}^{a_3 p_\rmt^{-1}} z^{M L_g - 1} e^{-\frac{z}{2}} \rmd z} \nonumber \\
	& = \lim_{p_\rmt \to \infty} \frac{-a_3^{ M L_g} p_\rmt^{\!-\! ML_g \!-\! 1} \exp\left(-\frac{1}{2} a_3 p_\rmt^{\!-\! 1}\right) \left(M L_g \!+\! \frac{1}{2} a_3 p_\rmt^{\!-\! 1}\right)}{-a_3^{M L_g} p_\rmt^{-ML_g - 1} \exp\left(-\frac{1}{2} a_3 p_\rmt^{-1}\right)} \nonumber \\
	& = M L_g.
\end{align}
where the first-order derivative is given by
\begin{align}
	P_{\rm out}'(p_\rmt) = -\frac{1}{2^{M L_g} \Gamma(M L_g)} a_3^{M L_g} p_\rmt^{-M L_g - 1} e^{-\frac{1}{2} a_3 p_\rmt^{-1}}.
\end{align}
From (\ref{diversity2}), it can be observed that both spatial diversity and frequency diversity are achievable in this scenario. In other words, for a deterministic forward-link channel, the BBBC system behaves similar to a conventional single-carrier block transmission system in Rayleigh fading, where each transmit antenna employs a transmit power of $\beta_1 p_\rmt$.

\section{Simulation Results} \label{simulationresults}

In this section, simulation results are presented to demonstrate the effectiveness of our proposed BBBC system. The default system parameters are listed in TABLE \ref{tab:system_parameters}.
Particularly, the large-scale fading coefficient is modeled as a function of distance, given by $\beta = 10^{-3} d^{-v}$, where $d$ and $v$ denote the distance and the path-loss exponent. The distances between the CE and the reader, between the CE and the BD, and between the BD and the reader are set to $d_\rmd = 100$ m, $d_f = 10$ m, and $d_g = 100$ m, respectively. The path-loss exponents are set to $v_\rmd = 3$, and $v_f = v_g = 2$ for the respective links. The small-scale fading in all channels follows the Rayleigh fading model. Moreover, the average SNR is defined as $p_\rmt \beta_1 \beta_2 / \sigma^2$. The following results are generated based on $10^5$ channel realizations.

\begin{table}[t!]
\caption{System Parameters} % 表格标题
\label{tab:system_parameters} % 用于引用的标签
\centering % 使表格居中显示
\begin{tabular}{|c|c|}
\hline
\textbf{Parameter} & \textbf{Value} \\
\hline
Block size, $N$ & $128$  \\
\hline
Length of ZP, $N_{\rm zp}$ & $16$  \\
\hline
Number of pilot symbol blocks, $N_\mathrm{p}$ & $2$  \\
\hline
Noise power, $\sigma^2$ & $-100$ dBm \\
\hline
Modulation scheme & QPSK \\
\hline
Number of antennas, $M$ & $16$ \\
\hline
Cyclic delay, $D$ & $\lfloor N / M\rfloor = 8$\\
\hline
Number of channel taps, $L_g$ & $4$ \\
\hline
\end{tabular}
\end{table}

\begin{figure}[t!]
    \centering    	
    \includegraphics[width=0.99\columnwidth]{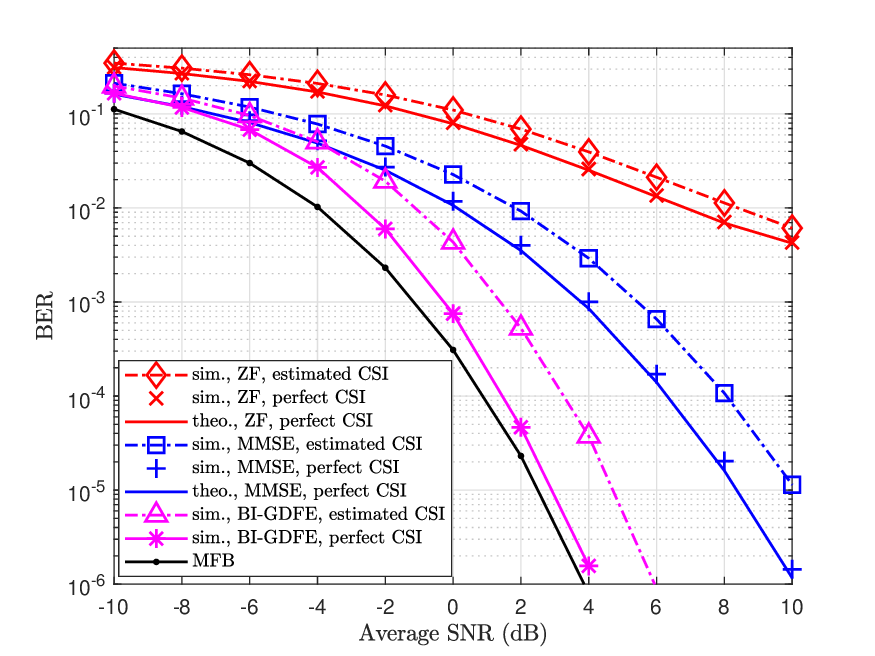}
    \caption{BER versus average SNR for perfect and estimated CSI.}	
    \label{BER_vs_SNR_perfect_estimated_CSI}	
    \vspace{-0.3cm}
\end{figure}

Fig.~\ref{BER_vs_SNR_perfect_estimated_CSI} depicts the BER performance versus the average SNR for different equalizers with perfect and estimated CSI. 
The theoretical BERs are obtained based on the SNR expressions in (\ref{snrzf}) and (\ref{snrmmse}).
First, it can be observed that the theoretical results for both ZF and MMSE equalizers align closely with the simulated results with perfect CSI. 
Furthermore, with the help of the proposed receiver design, the BER curves for estimated CSI show similar trends as those for perfect CSI.
For the considered equalizers, the gap between the BER performance for perfect and estimated CSI is below $2$ dB. This indicates that the proposed interference mitigation method effectively reduces direct-link interference. Moreover, the channels can be estimated with the proposed estimation method and pilot design. Furthermore, due to the noise amplification of the ZF equalizer, the slope of its BER curve is limited by $1$. In contrast, since the MMSE equalizer effectively balances ISI mitigation and noise amplification, the slope of its BER curve is much larger than $1$ in the high SNR regime, and is close to the slope of the MFB. Meanwhile, the BER performance of the BI-GDFE approaches that of the MFB in the high SNR regime. This demonstrates that the proposed system can achieve transmit diversity using existing equalizers like MMSE and BI-GDFE.

\begin{figure}[t!]
    \centering 
    \begin{subfigure}{.99\columnwidth}
        \centering
        \includegraphics[width=0.99\columnwidth]{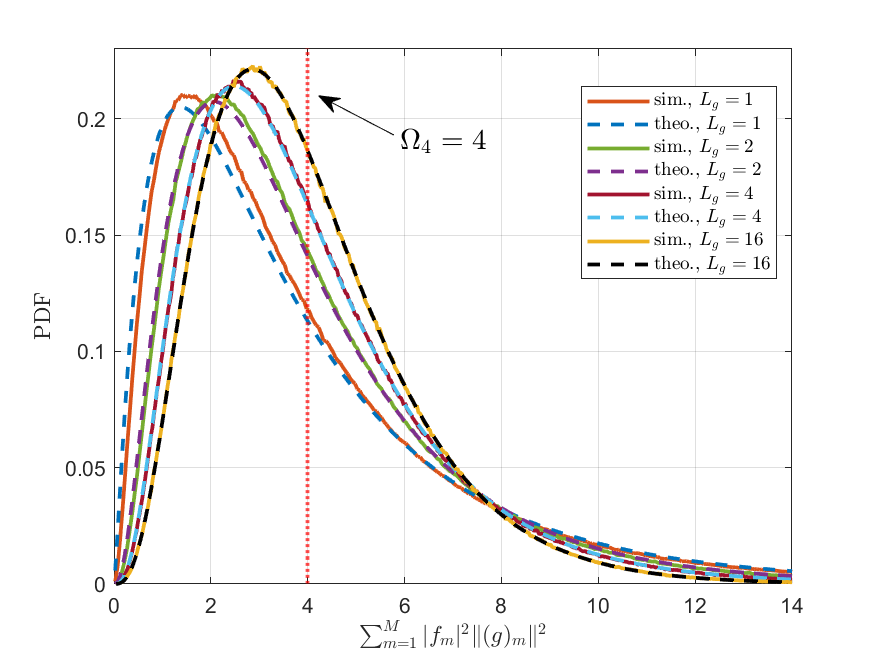}
        \caption{PDF for fixed $M = 4$.}
        \label{PDF_vs_L}	
    \end{subfigure}
    \begin{subfigure}{.99\columnwidth}
        \centering    	
        \includegraphics[width=0.99\columnwidth]{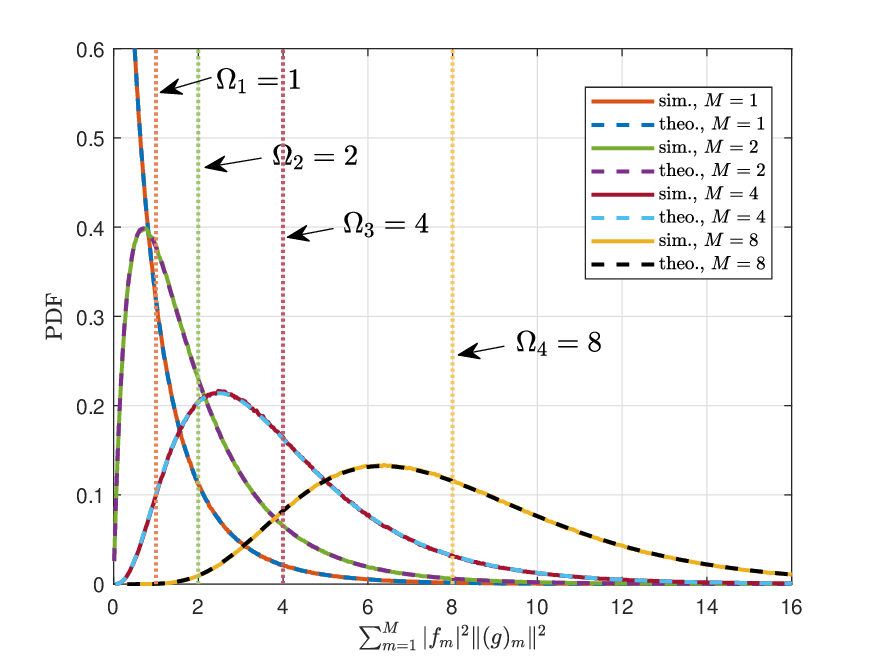}
        \caption{PDF for fixed $L_g = 4$.}	
        \label{PDF_vs_M}
    \end{subfigure}
    \caption{PDF of the equivalent channel gain under the general channel conditions.}
    \label{PDF}
    \vspace{-0.3cm}
\end{figure}

Fig.~\ref{PDF} depicts the PDF of the equivalent channel gain $\sum_{m = 1}^{M}|f_m|^2 \|\bg_m\|^2$ under the general channel conditions, with the path loss of both the forward and backward links normalized to $1$. 
The theoretical results are obtained by substituting the parameters from (\ref{parameters}) into the PDF expression in (\ref{kgpdf}). Specifically, Fig.~\ref{PDF_vs_L} focuses on scenarios with the number of antennas fixed at $M = 4$. First, the theoretical and simulated results align closely, especially for $L_g \ge 4$. Moreover, the increase in $L_g$ reduces the variance of the equivalent channel gain, leading to a narrower peak of the PDF that shifts toward the mean value of the channel gain $\Omega_4 = 4$. This indicates that an increasing $L_g$ can result in enhanced stability of the equivalent channel gain, benefiting backscatter communications. In addition, Fig.~\ref{PDF_vs_M} considers scenarios where the number of channel taps is fixed at $L_g = 4$. The results show that as $M$ increases, the equivalent channel gain can be significantly enhanced due to the increase in its mean value. Correspondingly, the array gain can also increase proportionally, confirming the advantages of deploying multiple antennas at the BD.

\begin{figure}[t!]
    \centering
    \begin{subfigure}{.99\columnwidth}
        \centering
        \includegraphics[width=0.99\columnwidth]{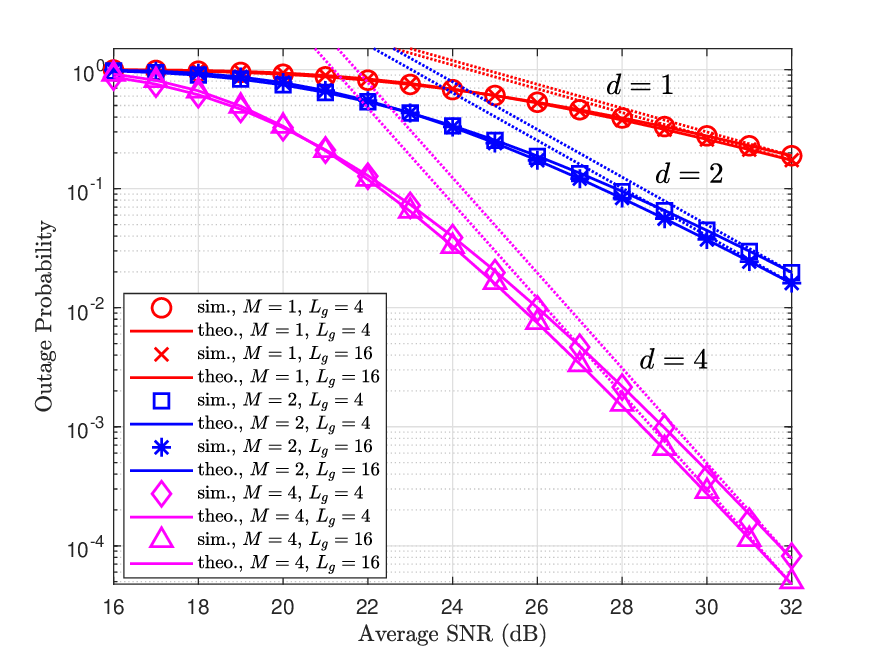}
	\caption{Under the general channel conditions.}
	\label{op_general_case}
    \end{subfigure}
    \begin{subfigure}{.99\columnwidth}
        \centering    
        \includegraphics[width=0.99\columnwidth]{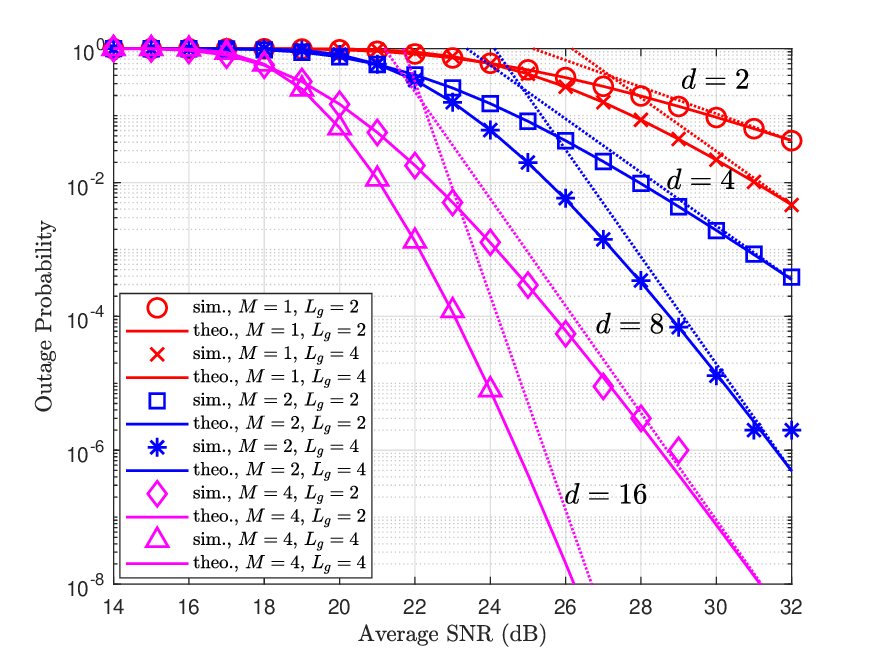}
        \caption{Under the special channel conditions.}
        \label{op_special_case}
    \end{subfigure}
    \caption{Outage probability versus average SNR.}
    \vspace{-0.3cm}
\end{figure}

\begin{figure}[t!]
    \centering
    \begin{subfigure}{.99\columnwidth}
        \centering    
        \includegraphics[width=0.99\columnwidth]{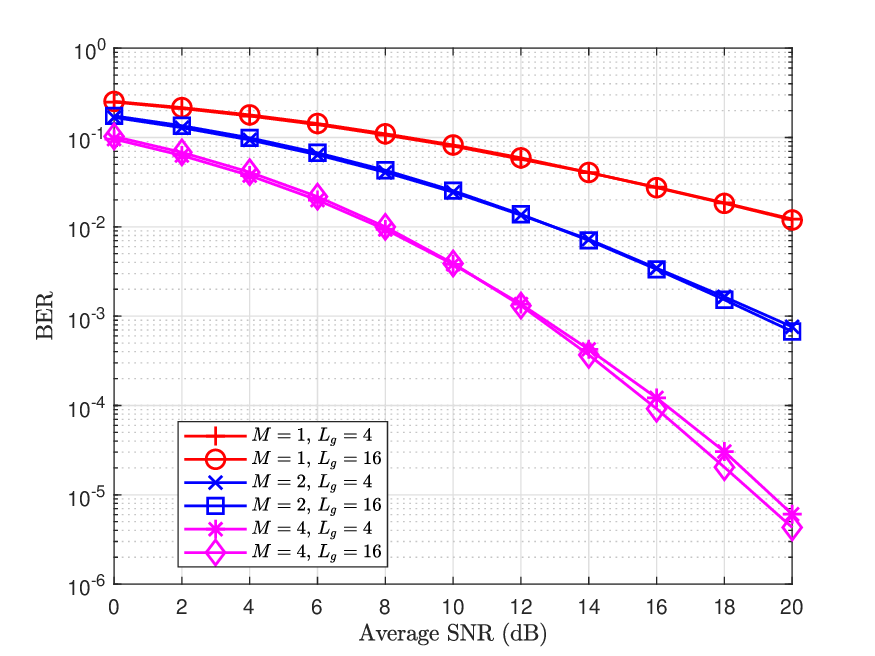}
        \caption{Under the general channel conditions.}
        \label{BER_general_case}
    \end{subfigure}
    \begin{subfigure}{.99\columnwidth}
        \centering    
        \includegraphics[width=0.99\columnwidth]{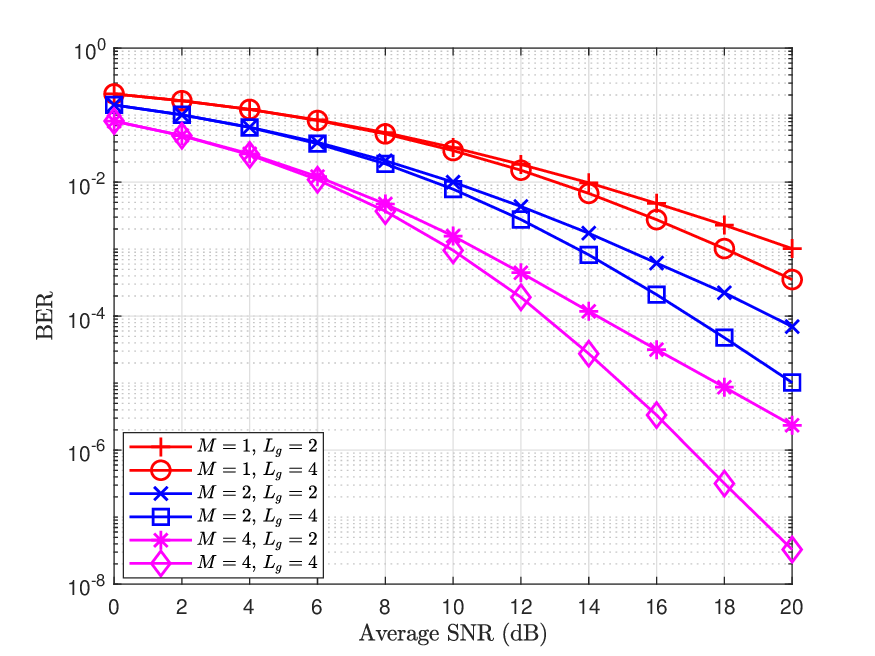}
        \caption{Under the special channel conditions.}
        \label{BER_special_case}
    \end{subfigure}
    \caption{BER versus average SNR with perfect CSI.}
    \vspace{-0.3cm}
\end{figure}

Fig.~\ref{op_general_case} depicts the outage probability versus the average SNR for varying values of $M$ and $L_g$ under the general channel conditions. The theoretical results are obtained from (\ref{op_expressions_general_case}). First, the overlap between the theoretical and simulated outage probabilities confirms the accuracy of our analysis. Notably, the slopes of the BER curves depend on the number of antennas $M$ instead of the number of channel taps $L_g$. Specifically, for $M = 1, 2,$ and $4$, the slopes in the high SNR regime are equal to $1$, $2$, and $4$, respectively. Furthermore, for $M = 4$ with $L_g = 4$ and $16$, the slopes are equal to $4$. This demonstrates that only spatial diversity can be exploited under the considered general channel conditions due to the channel fading in the forward link. However, increasing $L_g$ reduces the outage probability, which results from the fact that increased frequency selectivity can improve the SNR of the backscattered signal.

Fig.~\ref{op_special_case} depicts the outage probability versus the average SNR for varying values of $M$ and $L_g$ under the special channel conditions. The theoretical results are obtained from (\ref{op_expression_special_case}). First, it can be observed that the theoretical results match the simulated ones well. Moreover, unlike for the general channel conditions, the slopes of the BER curves for the special channel conditions are determined by the product of the number of antennas $M$ and the number of channel taps $L_g$. For example, when $M = 4$ and $L_g = 2$, the corresponding slope is $8$ in the high SNR regime. This indicates that both frequency and spatial diversity are attainable for the considered special channel conditions thanks to the constant input signal power at the BD.

Furthermore, to validate the consistency of the diversity order in terms of BER and outage probability, Fig.~\ref{BER_general_case} and Fig.~\ref{BER_special_case} depict the BER versus the average SNR for perfect CSI under the general and special channel conditions, respectively. The BER curves, generated using the MMSE equalizer rather than the ideal MFB, suffer from a performance loss compared with the MFB. However, the slopes of the BER curves closely align with those of the outage probability curves depicted in  Fig.~\ref{op_general_case} and Fig.~\ref{op_special_case}, respectively, confirming equal diversity performance.

\begin{figure}[t!]
	\centering    
	\includegraphics[width=0.99\columnwidth]{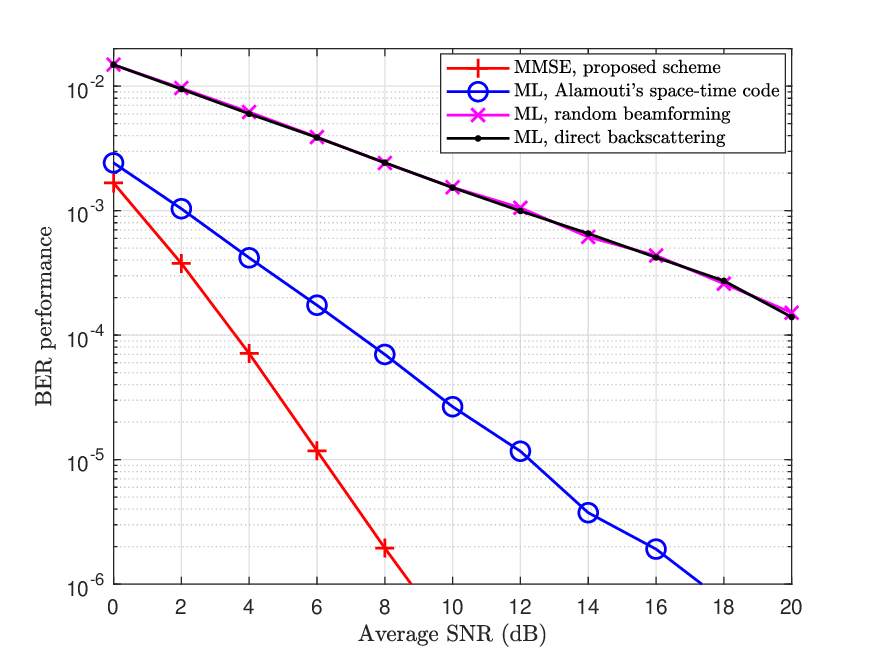}
	\caption{BER versus average SNR with $M = 32$ and $L_g = 1$.}
	\label{BER_vs_snr_comparison}
	\vspace{-0.3cm}
\end{figure}

Fig.~\ref{BER_vs_snr_comparison} depicts the BER versus the average SNR for different transmission schemes that do not require CSI. Flat fading is assumed for the backward link, i.e., $L_g = 1$. Meanwhile, the number of BD antennas is fixed at $M = 32$. We evaluate several benchmark schemes, including Alamouti's space-time code, random beamforming, and direct backscattering. Specifically, for Alamouti's code, BD antennas are divided into two groups to transmit the encoded signal. Within each group, the antennas send identical signals. For random beamforming and direct backscattering, the beamforming vector is randomly generated and fixed as an all-one vector, respectively.
% For spatial modulation, each pair of bits is divided such that one bit determines the antenna group while the other one modulates a binary PSK (BPSK) symbol backscattered by that group of antennas. 
Maximum likelihood (ML) detection is used by the benchmark schemes, while the MMSE equalizer is used by the proposed scheme to 
reduce the computational complexity. We notice that within the considered range of average SNR, the proposed scheme significantly outperforms the benchmark schemes. In particular, with the MMSE equalizer, the slope of the BER curve for the proposed scheme is much steeper than those of the benchmark schemes using ML detection, which have slopes of $1$ for random beamforming and direct backscattering, and $2$ for Alamouti's code. This enhanced performance is due to the effective utilization of multiple BD antennas in the proposed scheme, which facilitates both increased array gain and greater spatial diversity. Meanwhile, linear equalizers, like MMSE, can easily extract the enhanced diversity gain from the received signal with the proposed scheme. Conversely, the benchmark schemes, despite using an ML detector and multiple BD antennas, are only capable of achieving an array gain and a restricted diversity gain.

\begin{figure}[t!]
	\centering    
	\includegraphics[width=0.99\columnwidth]{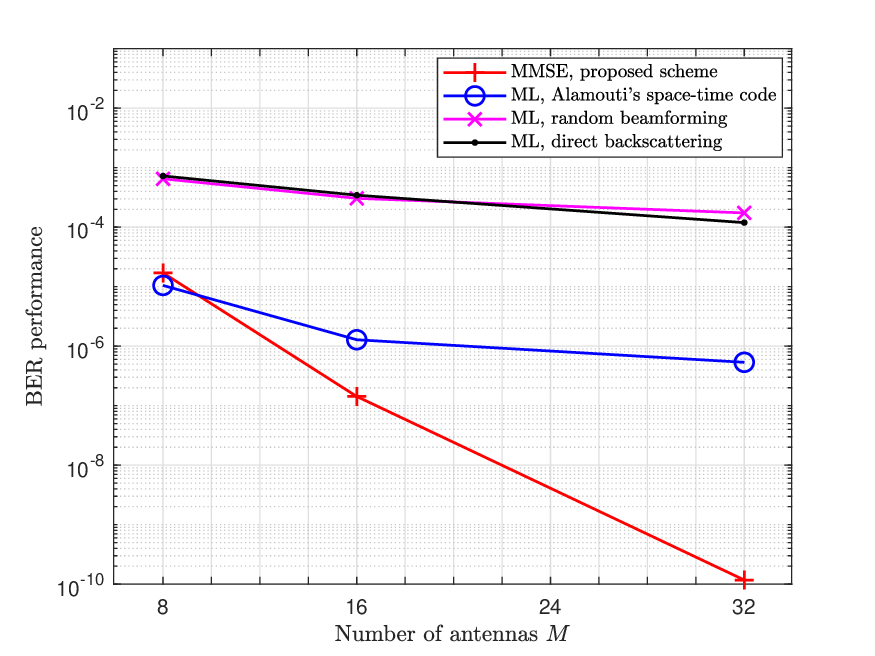}
	\caption{BER versus the number of antennas with $L_g = 1$.}
	\label{BER_comparison}
	\vspace{-0.3cm}
\end{figure}

Fig.~\ref{BER_comparison} depicts the BER versus the number of BD antennas $M$ for different transmission schemes. The average SNR is fixed at $20$ dB. It can be observed that the proposed scheme outperforms the benchmark schemes, especially for a larger $M$. Benefiting from the low-complexity decoding, ML detection allows Alamouti's code to outperform the proposed scheme employing MMSE equalization when $M < 10$. However, thanks to the enhanced spatial diversity for the proposed scheme, BER decreases as the number of antennas increases, whereas the benchmark schemes exhibit only slight improvements attributed to a growing array gain. As a result, the proposed scheme outperforms Alamouti's code for $M\ge 10$. Moreover, the proposed scheme can be employed for the case of $L_g > 1$, while Alamouti's code cannot straightforwardly be employed to address frequency-selective fading channels.

\section{Conclusion} \label{conclusion}

In this paper, we proposed a BBBC system with a multi-antenna BD. To overcome the frequency-selective fading channel in the backscattered link, we employed a single carrier block transmission scheme and implemented CDD at the BD to improve transmission reliability. Furthermore, we developed a receiver design based on frequency-domain equalization to efficiently recover the backscattered signal. Considering two different types of channel conditions, the outage probability and diversity order were analyzed for the proposed system. For the general channel conditions, where the forward and backward links both experience Rayleigh fading, spatial diversity is achievable. For the special channel conditions, where the forward-link channel is a deterministic LoS channel, both frequency and spatial diversity can be realized. Finally, simulation results validate the effectiveness of our proposed system under frequency-selective fading. Moreover, our results confirm that the proposed scheme can achieve both an improved array gain and an enhanced diversity order, and thereby improve the reliability of backscatter communications. Hence, the proposed BBBC system can be considered a promising candidate for backscatter communications with improved data rate and larger communication range.

\bibliography{IEEEabrv,reference}

\bibliographystyle{IEEEtran}

\end{document}